\documentclass[11pt]{article}
\usepackage{moriond}
\usepackage{graphicx}

\begin{document}
\hfill MZ-TH/09-14\\
\strut\hfill HU-EP-09/21\\ 
\strut\hfill SFB-CPP-09-40\\
\vspace*{2cm}
\title{Phenomenological studies of top-pair production + jet at NLO}

\author{S. Dittmaier$^{1,2}$, P. Uwer$^3$ and S. Weinzierl$^4$}

\address{
$^1$ Physikalisches Institut, Universit\"at Freiburg, D-79104 Freiburg, Germany
\\
$^2$ Max-Planck-Institut f\"ur Physik (Werner-Heisenberg-Institut), D-80805 M\"unchen, Germany
\\
$^3$ Institut f\"ur Physik,
Humboldt-Universit\"at zu Berlin, D-12489 Berlin, Germany
\\
$^4$ Institut f\"ur Physik, Universit\"at Mainz, D-55099 Mainz, Germany
}

\maketitle\abstracts{
In this talk we discuss top-quark pair production in association with an additional jet.
We present numerical results based on a next-to-leading order QCD calculation.
}

\section{Introduction}

Top-quark physics is currently studied at the Tevatron and will be on the agenda of the two main
LHC experiments, as soon as the LHC is turned on.
The top-quark is by far the heaviest elementary fermion in the 
Standard Model.
The large top mass is close to the scale of electroweak symmetry breaking and
it is reasonable to expect that the top-quark is particular sensitive to the details of the mechanism 
of electroweak symmetry breaking.

The production of a top-quark pair together with an additional jet is
an important reaction. This is clear from the simple
observation that a substantial number of events in the inclusive 
top-quark sample is accompanied by an additional jet. Depending on the
energy of the additional jet the fraction of events with an additional
jet can easily be of the order of 10--30\% or even more. For example
at the LHC we find a cross section of 376~pb for the production of a 
top--antitop-quark pair with an additional jet with a transverse momentum
above 50~GeV. This is almost half of the total top-quark pair cross
section which is 806~pb~\cite{Moch:2008qy} if evaluated in 
next-to-leading order (NLO).

Ignoring the Standard Model as {\it the theory} of particle physics one might 
wonder whether the top-quark, which is almost as heavy as a gold
atom, behaves as a point-like particle. 
A deviation from the point-like nature would appear as anomalous 
moments yielding differential distributions different from
the point-like case. Anomalous couplings to the gluon are most
naturally probed via the production of an additional jet. 

The emission of an additional
gluon also leads to a rather interesting property of the cross
section: The differential cross section contains contributions from
the interference of C-odd and C-even parts of the 
amplitude$^{2-5}$,
where C denotes the charge conjugation.
While for the total cross section these
contributions cancel when integrating over the (symmetric) phase space
they can lead to a forward--backward charge 
asymmetry of the top-quark which is currently measured
at the Tevatron \cite{:2007qb,Aaltonen:2008hc}.

Apart from its
significance as signal process it turns out that $\mathrm{t} \bar{\mathrm{t}} + \mbox{jet}$ 
production is also an important background to 
various new physics searches. A prominent example is Higgs production
via vector-boson fusion. 
The major background to this reaction is due to 
$\mathrm{t} \bar{\mathrm{t}} + \mbox{jet}$, again underlining
the need for precise theoretical predictions for this process.
In this talk we discuss predictions based on an NLO calculation \cite{Dittmaier:2007wz,Dittmaier:2008uj}.

\section{The calculation}

As with any NLO calculation there are real and virtual corrections.
The matrix elements corresponding to the real emission contribution
are given by the square of the Born amplitudes with $6$ partons.
All relevant matrix elements for this contribution can be obtained 
by crossing from the generic matrix elements
\begin{equation}
0 \to \mathrm{t} \bar{\mathrm{t}} g g g g, 
\quad
0 \to \mathrm{t} \bar{\mathrm{t}} q \bar q g g,
\quad
0 \to \mathrm{t} \bar{\mathrm{t}} q \bar q q' \bar q',
\quad
0 \to \mathrm{t} \bar{\mathrm{t}} q \bar q q \bar q.
\end{equation}
The matrix elements for the virtual contribution
are given by the interference term
of the one-loop amplitudes with $5$ partons with 
the corresponding Born amplitude.
The required generic matrix elements are
\begin{equation}
0 \to \mathrm{t} \bar{\mathrm{t}} g g g, \quad 
0 \to \mathrm{t} \bar{\mathrm{t}} q \bar q g.
\end{equation}
Taken individually, the real and the virtual contributions are infrared divergent. Only their sum is finite.
We use the dipole subtraction 
formalism$^{10-12}$
to render the individual contributions finite.
The virtual one-loop diagrams consist of 
self-energy, vertex, box-type, and pentagon-type corrections.
The most complicated diagrams are the pentagon diagrams.
We use highly automated procedures for all contributions and various procedures to ensure numerically stability and 
efficiency$^{13-24}$.
To ensure correctness we have for every piece two independent implementations, using where ever possible different methods and 
tools$^{25-31}$.

\section{Numerical results}

For the numerical analysis we use the value $m_{\mathrm{t}}=174\mbox{GeV}$ for the top-quark mass, 
CTEQ6L1 PDFs together with a one-loop running $\alpha_{\mathrm{s}}$ in LO, 
as well as CTEQ6M PDFs with a two-loop running $\alpha_{\mathrm{s}}$ in NLO.
The number of active flavours is $N_{\mathrm{F}}=5$, and the
respective QCD parameters are $\Lambda_5^{\mathrm{LO}}=165\mbox{MeV}$
and $\Lambda_5^{\overline{\mathrm{MS}}}=226\mbox{MeV}$.
We identify the renormalisation and factorisation scales, $\mu=\mu_{\mathrm{ren}}=\mu_{\mathrm{fact}}$.
The additional jet is defined by the Ellis-Soper $k_\perp$-jet algorithm\cite{Ellis:1993tq} with $R=1$  and
we require a transverse momentum of
$p_{\mathrm{T,jet}}>p_{\mathrm{T,jet,cut}}$.
For the Tevatron we use $p_{\mathrm{T,jet,cut}}=20\mbox{GeV}$, for the LHC we either use
$p_{\mathrm{T,jet,cut}}=20\mbox{GeV}$ or $p_{\mathrm{T,jet,cut}}=50\mbox{GeV}$.
\begin{figure}
\begin{center}
\includegraphics[bb= 202 433 458 655,width=0.3\textwidth]{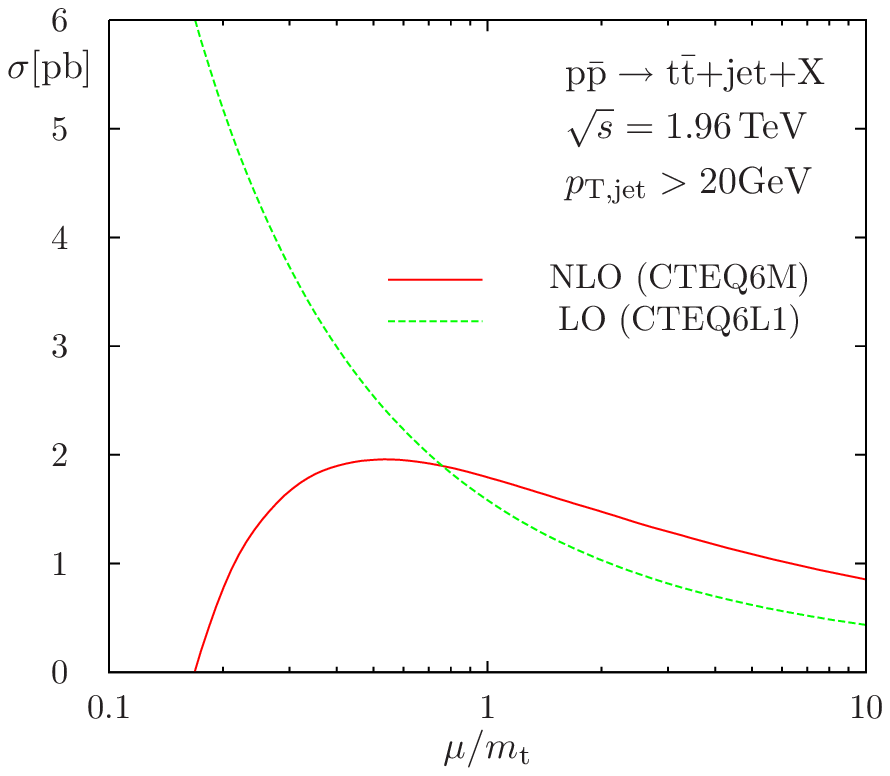}
\includegraphics[bb= 202 433 458 655,width=0.3\textwidth]{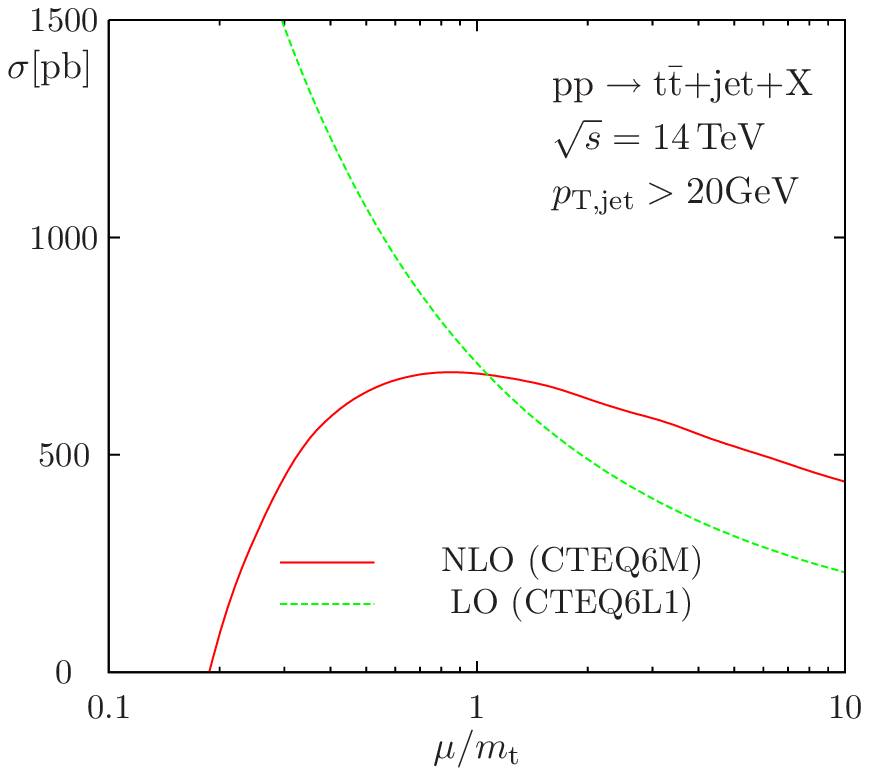}
\includegraphics[bb= 222 458 458 655,width=0.3\textwidth]{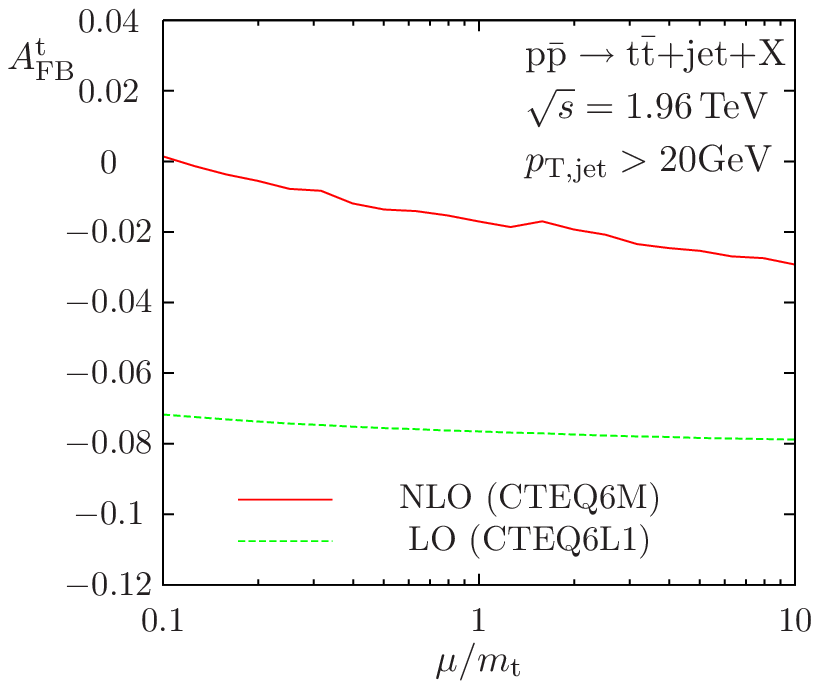}
\end{center}
\caption{
\label{fig_1}
Scale dependence at LO (green) and NLO (red).
The left plot shows the cross section for $\mathrm{t} \bar{\mathrm{t}}{+}$jet production at the Tevatron,
the middle one shows the corresponding plot for the LHC.
The right plot shows the forward-backward charge asymmetry at the Tevatron.
}
\end{figure}
Fig.~\ref{fig_1} shows the scale dependence of the cross section at LO and NLO for the Tevatron and the LHC.
The NLO corrections reduce the scale dependence significantly.
In addition we show in Fig.~\ref{fig_1} the forward-backward charge asymmetry at the Tevatron.
In LO the asymmetry is defined by
\begin{equation}
 A^{\mathrm{t}}_{\mathrm{FB,LO}} = 
\frac{\sigma^-_{\mathrm{LO}}}{\sigma^+_{\mathrm{LO}}},
\quad
\sigma^\pm_{\mathrm{LO}} = 
\sigma_{\mathrm{LO}}(y_{\mathrm{t}}{>}0)\pm\sigma_{\mathrm{LO}}(y_{\mathrm{t}}{<}0),
\end{equation}
where $y_{\mathrm{t}}$ denotes the rapidity of the top-quark.
Denoting the corresponding NLO contributions to the cross sections by
$\delta\sigma^\pm_{\mathrm{NLO}}$,
we define the asymmetry at NLO by
\begin{equation}
 A^{\mathrm{t}}_{\mathrm{FB,NLO}} = 
\frac{\sigma^-_{\mathrm{LO}}}{\sigma^+_{\mathrm{LO}}}
\left( 1+
 \frac{\delta\sigma^-_{\mathrm{NLO}}}{\sigma^-_{\mathrm{LO}}}
-\frac{\delta\sigma^+_{\mathrm{NLO}}}{\sigma^+_{\mathrm{LO}}} \right),
\end{equation}
i.e.\ via a consistent expansion in $\alpha_{\mathrm{s}}$.
At LO we find an asymmetry of about
$-8$\%. The scale dependence is rather small. This is a consequence of
the fact that
$\alpha_s$ cancels exactly between the numerator and the
denominator. In addition the residual factorisation scale dependence
also cancels to a large extent in the ratio. 
The NLO corrections to the asymmetry are of order $\alpha_s^1$ 
and depend on the renormalisation scale.
\begin{figure}
\begin{center}
\includegraphics[bb= 160 407 424 714,width=0.24\textwidth]{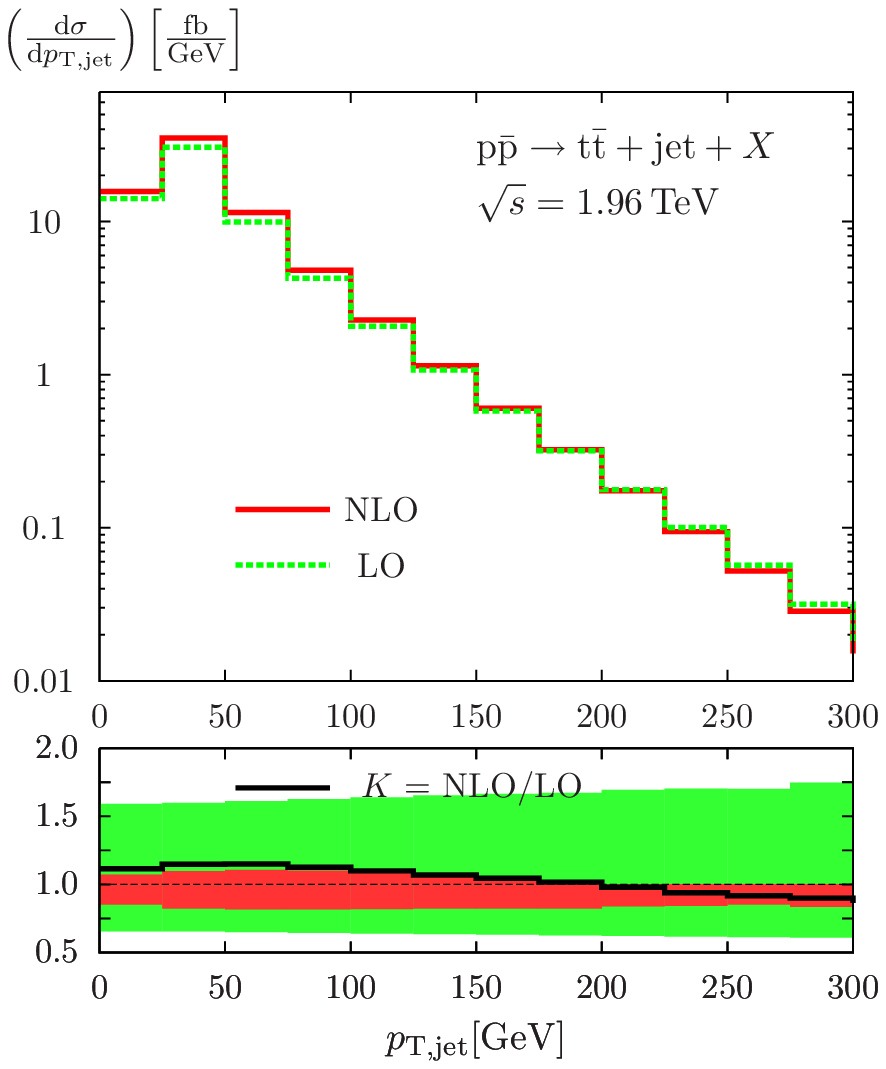}
\includegraphics[bb= 160 407 424 714,width=0.24\textwidth]{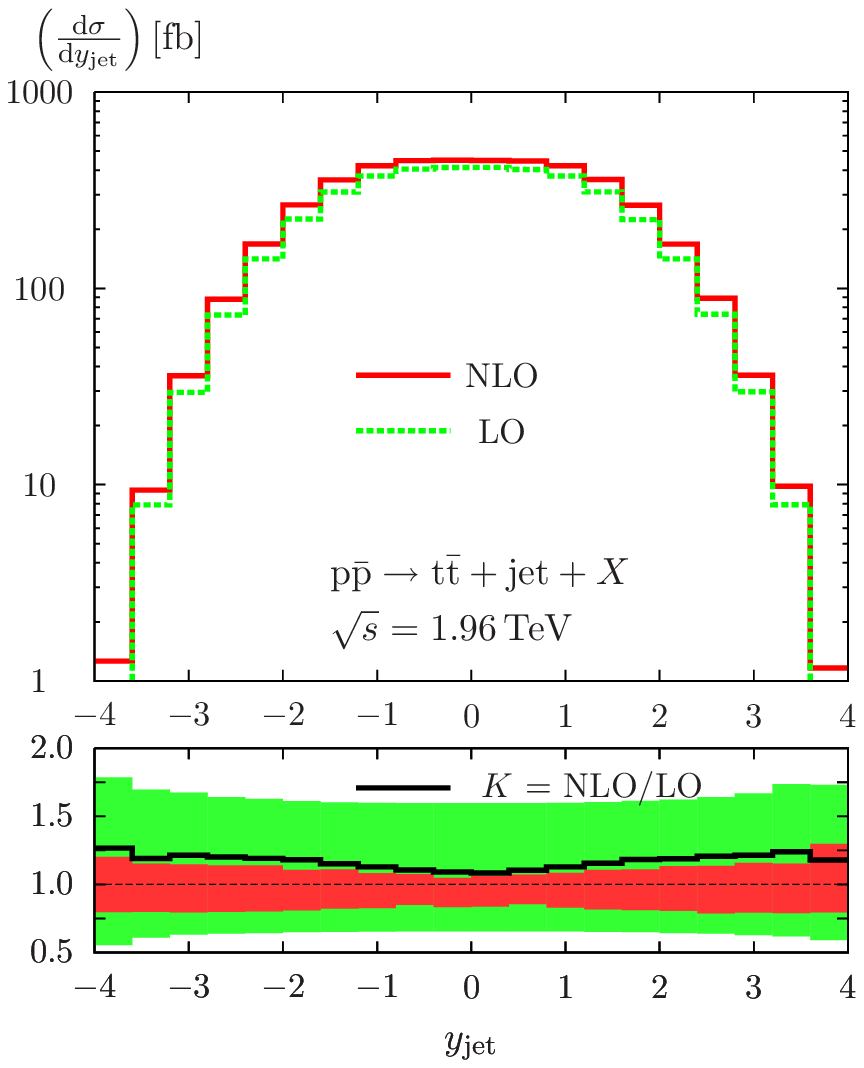}
\includegraphics[bb= 160 407 424 714,width=0.24\textwidth]{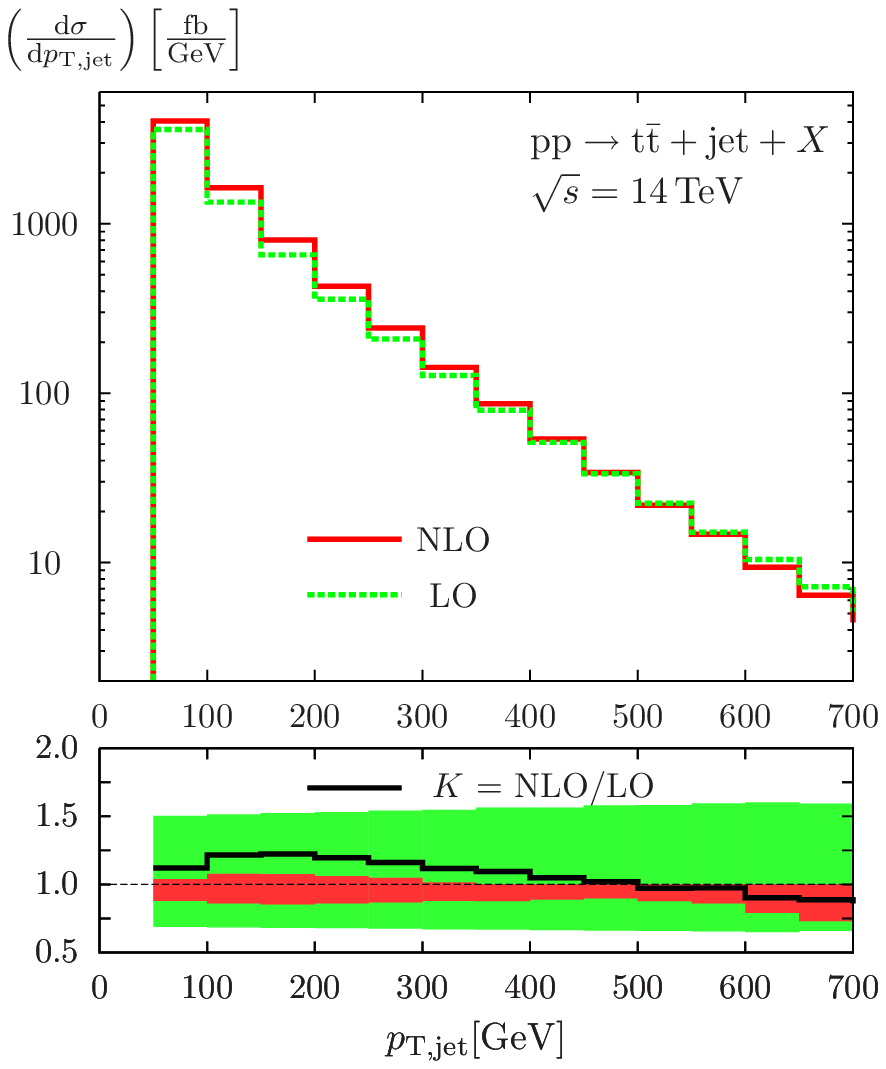}
\includegraphics[bb= 160 407 424 714,width=0.24\textwidth]{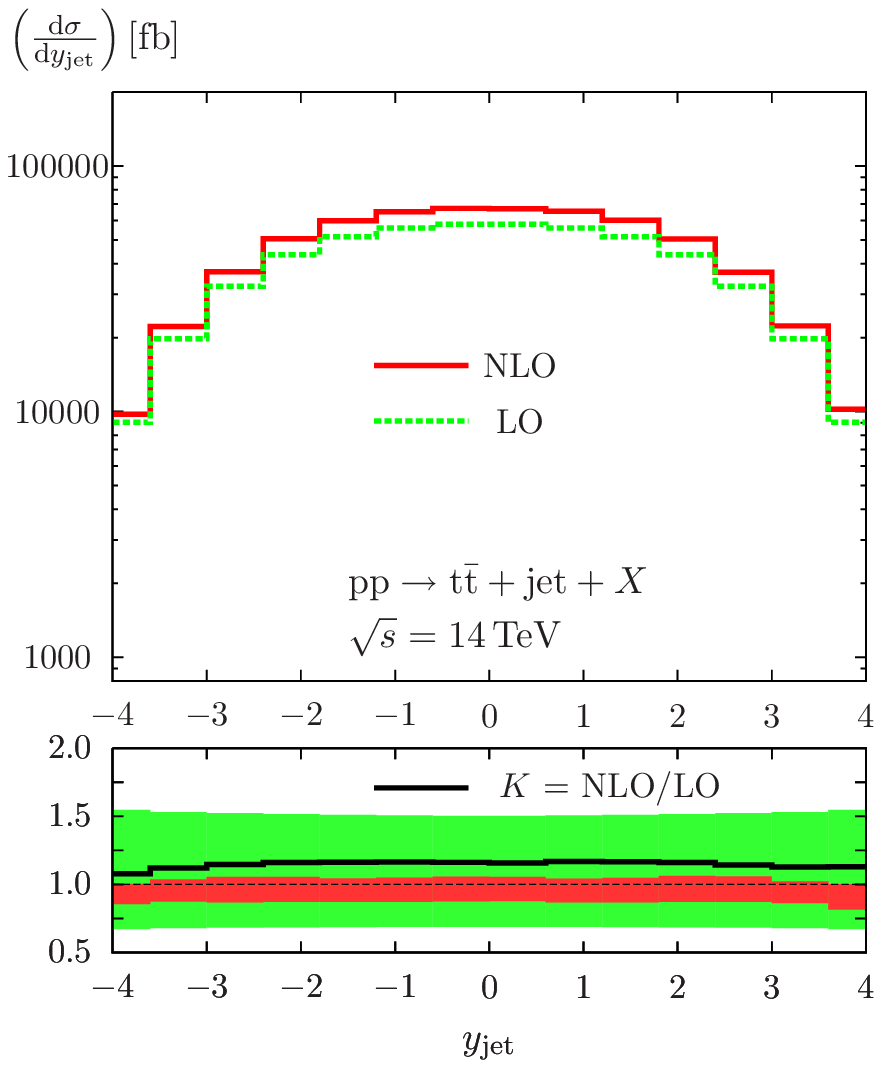}
\end{center}
\caption{
\label{fig_2}
The distribution of the additional jet in $p_{\mathrm{T}}$ and rapidity at the Tevatron (left) and the LHC (right).
The large plots show the predictions at LO (green) and NLO (red). The lower panels show the
scale uncertainties, the LO scale uncertainty is shown by a green band, the NLO scale uncertainty by
a red band. 
In addition the ratios $K=\mathrm{NLO/LO}$ are shown by a black line.
}
\end{figure}
It is therefore natural to expect a stronger scale 
dependence of the asymmetry at NLO than at LO, as seen in the plot.
Within the current numerical set-up the asymmetry is almost washed out at NLO.

In Fig.~\ref{fig_2} we show the distribution of the additional jet in $p_{\mathrm{T}}$ and rapidity at the Tevatron and the LHC.
In these plots we used $p_{\mathrm{T,jet,cut}}=20\mbox{GeV}$ for the Tevatron and $p_{\mathrm{T,jet,cut}}=50\mbox{GeV}$ for the LHC.
The scale variations in the lower panel correspond to a variation by a factor of $2$ around $\mu=m_{\mathrm{t}}$.
\begin{figure}
\begin{center}
\includegraphics[bb= 160 407 424 714,width=0.24\textwidth]{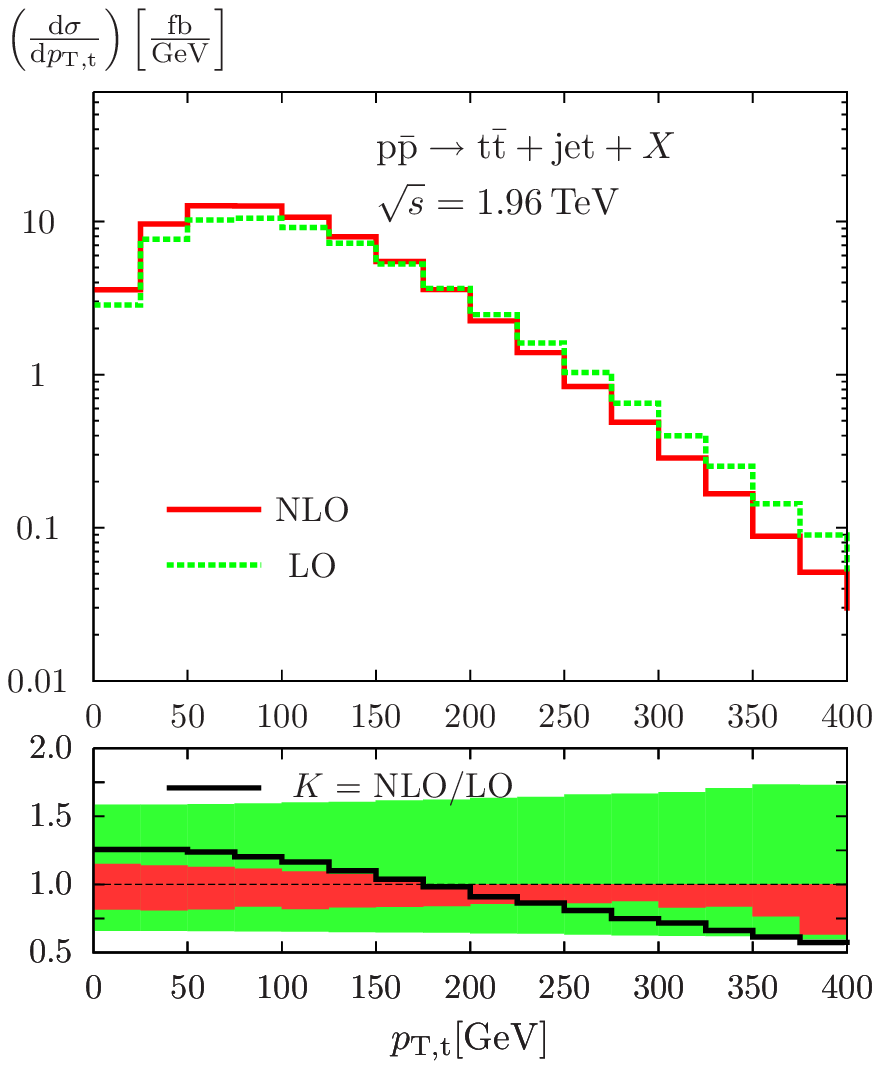}
\includegraphics[bb= 160 407 424 714,width=0.24\textwidth]{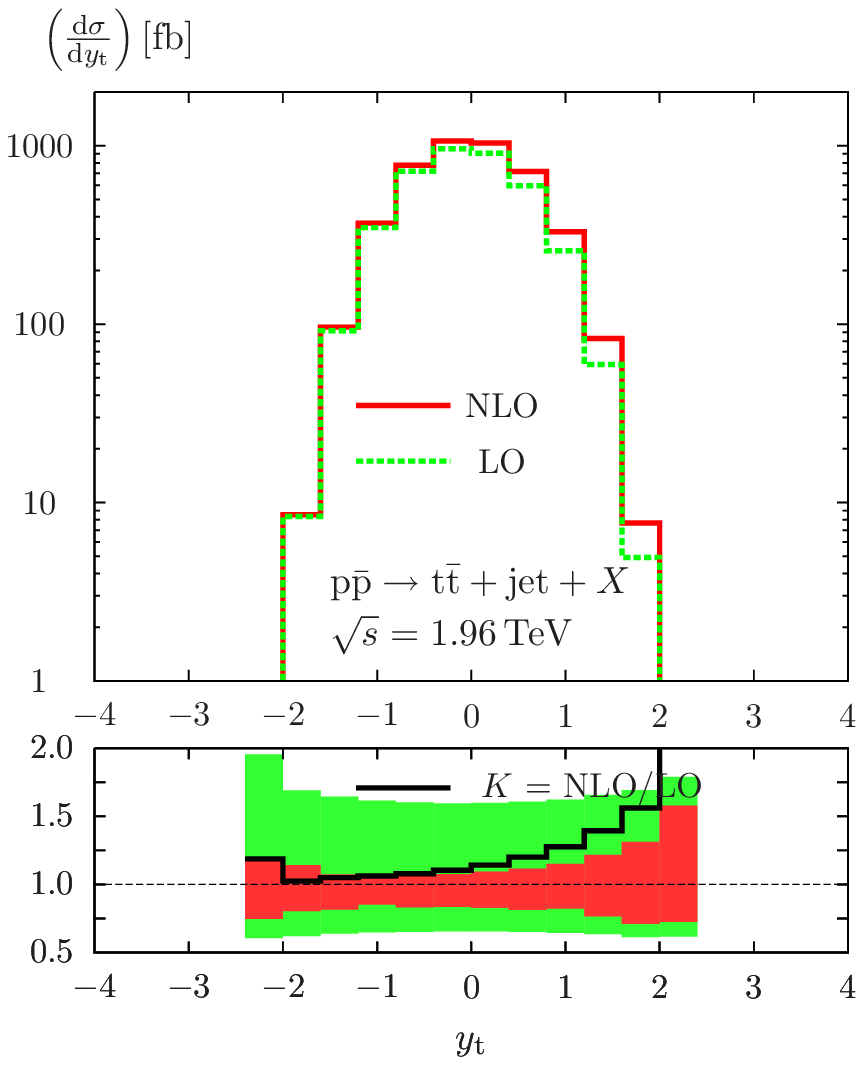}
\end{center}
\caption{
\label{fig_3}
The distribution of the top-quark in $p_{\mathrm{T}}$ (left) and rapidity (right) at the Tevatron.
}
\end{figure}
As expected, the scale uncertainties are reduced by the inclusion of the NLO corrections.
Also shown in the lower panel of Fig.~\ref{fig_2} is the ratio $K=\mathrm{NLO/LO}$.

Finally, Fig.~\ref{fig_3} shows the distribution of the top-quark in $p_{\mathrm{T}}$ and rapidity at the Tevatron.
We note that the NLO corrections do not simply rescale the LO shape, but induce distortions 
of the distributions, as can be seen from the non-constant $K$-factor.

\section{Conclusions}

In this talk we discussed 
predictions for $\mathrm{t} \bar{\mathrm{t}}{+}$jet production at hadron
colliders based on a NLO QCD calculation.
For the cross section the NLO corrections
reduce significantly the scale dependence of the LO predictions.
The charge asymmetry of the top-quarks at the Tevatron is significantly decreased at NLO and is almost washed out
in particular when the residual scale dependence is taken into account. 
Further refinements of the precise definition of the forward-backward asymmetry are required
to stabilise the  asymmetry with respect to higher-order corrections.
From a technical perspective the calculation we reported on represents a corner-stone for NLO multi-leg
computations, which are required for the LHC.


\section*{References}

\bibliography{moriond09}
\bibliographystyle{/home/stefanw/latex-style/h-physrev3}

\end{document}